\documentclass[twocolumn,times,tighten,astrosym]{aastex61}

\usepackage{graphicx, latexsym, graphicx, amssymb, longtable, epsf}
\usepackage{amssymb, amsmath, mathrsfs, subfigure, threeparttable}

\hypersetup{linkcolor=cyan, citecolor=cyan, filecolor=cyan, urlcolor=blue}

\newcommand{\myr}{${\rm M_{\sun}\,yr^{-1}}$}
\newcommand{\Msol}{${\rm M_{\sun}}$}

\newcommand{\um}{$\mu$m}

\newcommand{\Jykms}{${\rm Jy\,km\,s^{-1}}$}
\newcommand{\kms}{${\rm km}\,{\rm s}^{-1}$}
\newcommand{\alphaUnits}{${(\rm M_{\sun}}\,\rm(K\,km\,s^{-1}\,pc^{-2})^{-1})$}
\newcommand{\Lco}{$L^{\prime}_{\rm CO}$}

\newcommand{\alphaCO}{${\alpha_{\rm CO}}$}
\newcommand{\asec}{$^{\prime\prime}$}

\newcommand{\subfigsz}{0.69}
\newcommand{\subfigsztwo}{1.0}

\received{2017 May 8}
\revised{2017 June 5}
\accepted{2017 June 6}

\shorttitle{ALMA CO (2--1) in $z\sim1.6$ Cluster Galaxies}
\shortauthors{Noble et al.}

\begin{document}

\title{ALMA Observations of Gas-Rich Galaxies in $\lowercase{z}\sim1.6$ Galaxy Clusters: Evidence for Higher Gas Fractions in High-Density Environments}

\correspondingauthor{A.G.~Noble}
\email{noble@mit.edu}

\author{A.G.~Noble}
\affiliation{Kavli Institute for Astrophysics and Space Research, Massachusetts Institute of Technology, 77 Massachusetts Avenue, Cambridge, MA 02139, USA}

\author{M.~McDonald}
\affiliation{Kavli Institute for Astrophysics and Space Research, Massachusetts Institute of Technology, 77 Massachusetts Avenue, Cambridge, MA 02139, USA}

\author{A.~Muzzin}
\affiliation{Department of Physics and Astronomy, York University, 4700 Keele Street, Toronto, ON MJ3 1P3 Canada}

\author{J.~Nantais}
\affiliation{Departamento de Ciencias F\'{i}sicas, Universidad Andres Bello, Fernandez Concha 700, Las Condes 7591538, Santiago, Regi\'{o}n Metropolitana, Chile}

\author{G.~Rudnick}
\affiliation{The University of Kansas, Department of Physics and Astronomy, 1251 Wescoe Hall Drive, Lawrence, KS 66045, USA}

\author{E.~van Kampen}
\affiliation{European Southern Observatory, Karl-Schwarzschild-Strasse 2 D-85748 Garching bei M\"{u}nchen, Germany}

\author{T.M.A.~Webb}
\affiliation{Department of Physics, McGill University, 3600 rue University, Montr\'{e}al, QC H3A 2T8, Canada}

\author{G.~Wilson}
\affiliation{Department of Physics and Astronomy, University of California, Riverside, CA 92521, USA}

\author{H.K.C.~Yee}
\affiliation{Department of Astronomy and Astrophysics, University of Toronto, 50 St. George Street, Toronto, ON M5S 3H4, Canada}

\author{K.~Boone}
\affiliation{Department of Physics, University of California Berkeley, 366 LeConte Hall, MC 7300, Berkeley, CA 94720-7300, USA}

\author{M.C.~Cooper}
\affiliation{Department of Physics and Astronomy, University of California, Irvine, 4129 Frederick Reines Hall, Irvine, CA 92697, USA}

\author{A.~DeGroot}
\affiliation{Department of Physics and Astronomy, University of California, Riverside, CA 92521, USA}

\author{A.~Delahaye}
\affiliation{Department of Physics, McGill University, 3600 rue University, Montr\'{e}al, Qu\'{e}bec H3A 2T8, Canada}

\author{R.~Demarco}
\affiliation{Departamento de Astronom\'{i}a, Universidad de Concepci\'{o}n, Casilla 160-C, Concepci\'{o}n, Regi\'{o}n del Biob\'{i}o, Chile}

\author{R.~Foltz}
\affiliation{Department of Physics and Astronomy, University of California, Riverside, CA 92521, USA}

\author{B.~Hayden}
\affiliation{Department of Physics, University of California Berkeley, 366 LeConte Hall MC 7300, Berkeley, CA 94720-7300, USA}
\affiliation{Physics Division, Lawrence Berkeley National Laboratory, 1 Cyclotron Road, Berkeley, CA 94720, USA}

\author{C.~Lidman}
\affiliation{Australian Astronomical Observatory, 105 Delhi Road, North Ryde, NSW 2113, Australia}

\author{A.~Manilla-Robles}
\affiliation{European Southern Observatory, Karl-Schwarzschild-Strasse 2 D-85748 Garching bei M\"{u}nchen, Germany}

\author{S.~Perlmutter}
\affiliation{Department of Physics, University of California Berkeley, 366 LeConte Hall MC 7300, Berkeley, CA 94720-7300, USA}
\affiliation{Physics Division, Lawrence Berkeley National Laboratory, 1 Cyclotron Road, Berkeley, CA 94720, USA}

\begin{abstract}

We present ALMA CO (2--1) detections in 11 gas-rich cluster galaxies at $z\sim1.6$, constituting the largest sample of molecular gas measurements in $z>1.5$ clusters to date.  The observations span three galaxy clusters, derived from the \textit{Spitzer} Adaptation of the Red-sequence Cluster Survey.  We augment the $>5\sigma$ detections of the CO (2--1) fluxes with multi-band photometry, yielding stellar masses and infrared-derived star formation rates, to place some of the first constraints on molecular gas properties in $z\sim1.6$ cluster environments.  We measure sizable gas  reservoirs of $0.5-2\times10^{11}$\Msol\ in these objects, with high gas fractions ($f_{\rm gas}$) and long depletion timescales ($\tau$), averaging 62\% and 1.4\,Gyr, respectively.  We compare our cluster galaxies to the scaling relations of the coeval field, in the context of how gas fractions and depletion timescales vary with respect to the star-forming main sequence.  We find that our cluster galaxies lie systematically off the field scaling relations at $z=1.6$  toward enhanced gas fractions, at a level of $\sim4\sigma$, but have consistent depletion timescales.  Exploiting CO detections in lower-redshift clusters from the literature, we investigate the evolution of the gas fraction in cluster galaxies, finding it to mimic the strong rise with redshift in the field.  We emphasize the utility of detecting abundant gas-rich galaxies in high-redshift clusters, deeming them as crucial laboratories for future statistical studies. 
\end{abstract}


\section{Introduction}

Galaxy cluster evolution is intertwined with that of its constituent galaxies.  Therefore, in order to understand the former, we must explore the baryonic processes that shape the latter.   In particular, this requires a solid understanding of the molecular gas content in cluster galaxies, as this provides the necessary raw material to fuel star formation.  Within these dense environments, cluster galaxies face hostile conditions, as substantiated by morphological and physical transformations.  Various mechanisms have been invoked to explain the differences between cluster and field galaxies, many of which involve interactions with the intracluster medium (ICM).  For example, ram-pressure stripping has been directly observed in low-redshift ($z\lesssim0.3$) cluster galaxies via H\textsc{i} deficiencies \citep{Jaffe15}, extraplanar H\textsc{i} gas \citep{Chung09}, and long ``jellyfish" gas tails \citep{Owers12}.

The environmental effect of the ICM on molecular gas, however, is more ambiguous.  It is thought that this denser gas is less susceptible to removal and can therefore survive the effects of ram-pressure stripping.  Indeed, many studies have found no difference in the molecular gas content between field and cluster environments, as traced by the emission lines of $^{12}$CO  \citep[e.g.,][]{Stark86, KenneyYoung89}. 
However, more recent work has reported molecular gas deficiencies in cluster galaxies  \citep[e.g.,][]{Fumagalli09, Jablonka13, Scott13, Boselli14}.

Technological advances in radio interferometers have enabled statistical samples of CO in the field out to $z\sim3$ \citep[e.g.][]{Saintonge11, Tacconi13, Tacconi17}.  Cluster samples, however, have primarily focused on low-redshift systems. A missing key component of molecular gas studies are observations within high-redshift cluster cores.

Observations  \citep[][]{Tran10, Brodwin13} suggest that $z\gtrsim1.5$ is the peak assembly time for galaxy clusters, and it is thus likely that many of the environmental effects on cluster galaxies occur at these early times in dense regions.
 While there have been some molecular gas observations in the dense regions of $z>1$ clusters, these have been limited to only a handful of  detections \cite[][Rudnick et al. ApJ submitted]{Aravena12, Wagg12, Casasola13, Hayashi17}.  Thus, whether high-redshift clusters typically harbor gas-rich galaxies, or whether they are analogous to some of their lower-redshift counterparts, displaying signs of molecular gas deficiencies, has yet to be determined conclusively.

Here, we present Cycle 3 ALMA observations of three massive galaxy clusters at $z\sim1.6$ from the \textit{Spitzer} Adaptation of the Red-sequence Cluster Survey (SpARCS).  With a total of 11 CO (2--1) detections at $>5\sigma$, we are filling in this CO redshift desert and enabling the first statistical constraints on gas properties in high-redshift cluster galaxies.  Stellar masses and star formation rates (SFRs) are based on a Chabrier initial mass function \citep{Chabrier03}.
	
 \begin{figure*}[] \centering
{\includegraphics[width=\subfigsz\columnwidth]{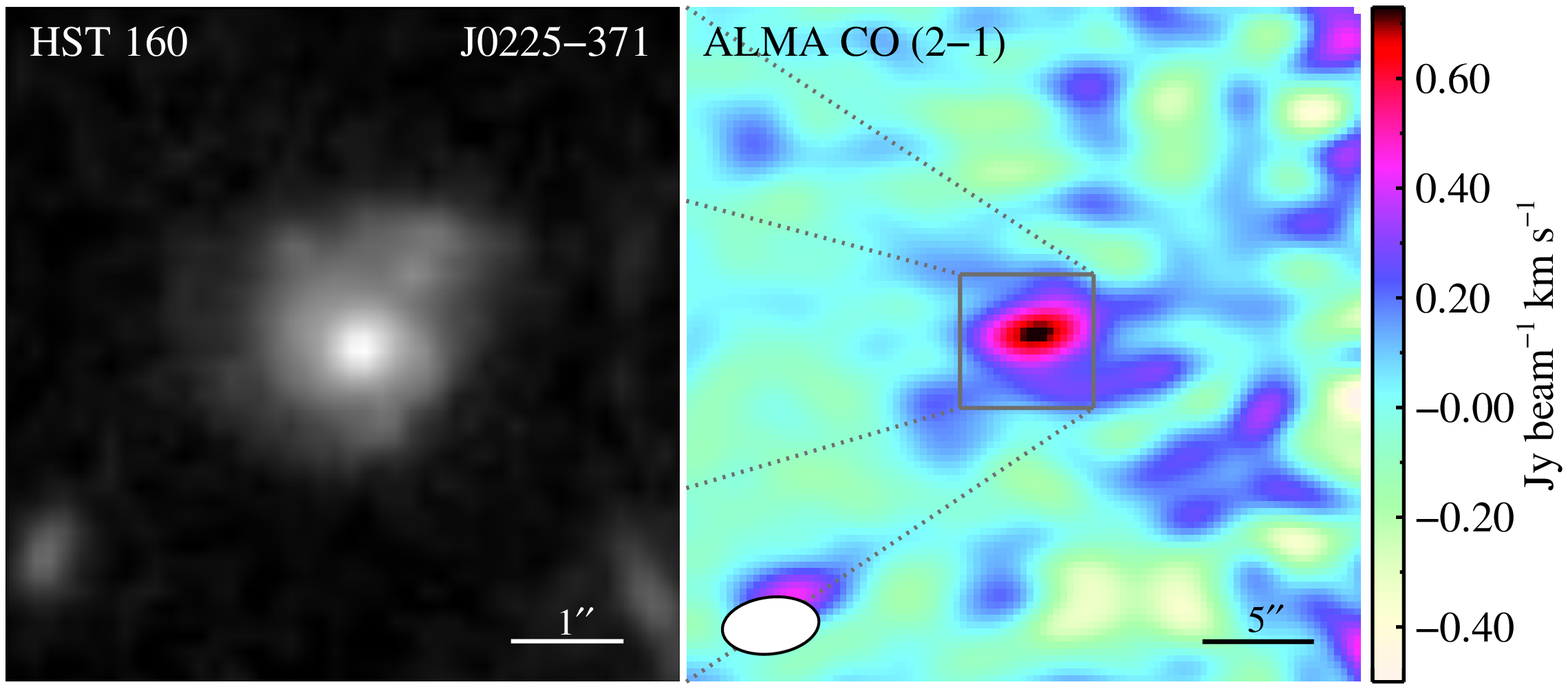}}\hfill%
\subfigure{\includegraphics[width=\subfigsz\columnwidth]{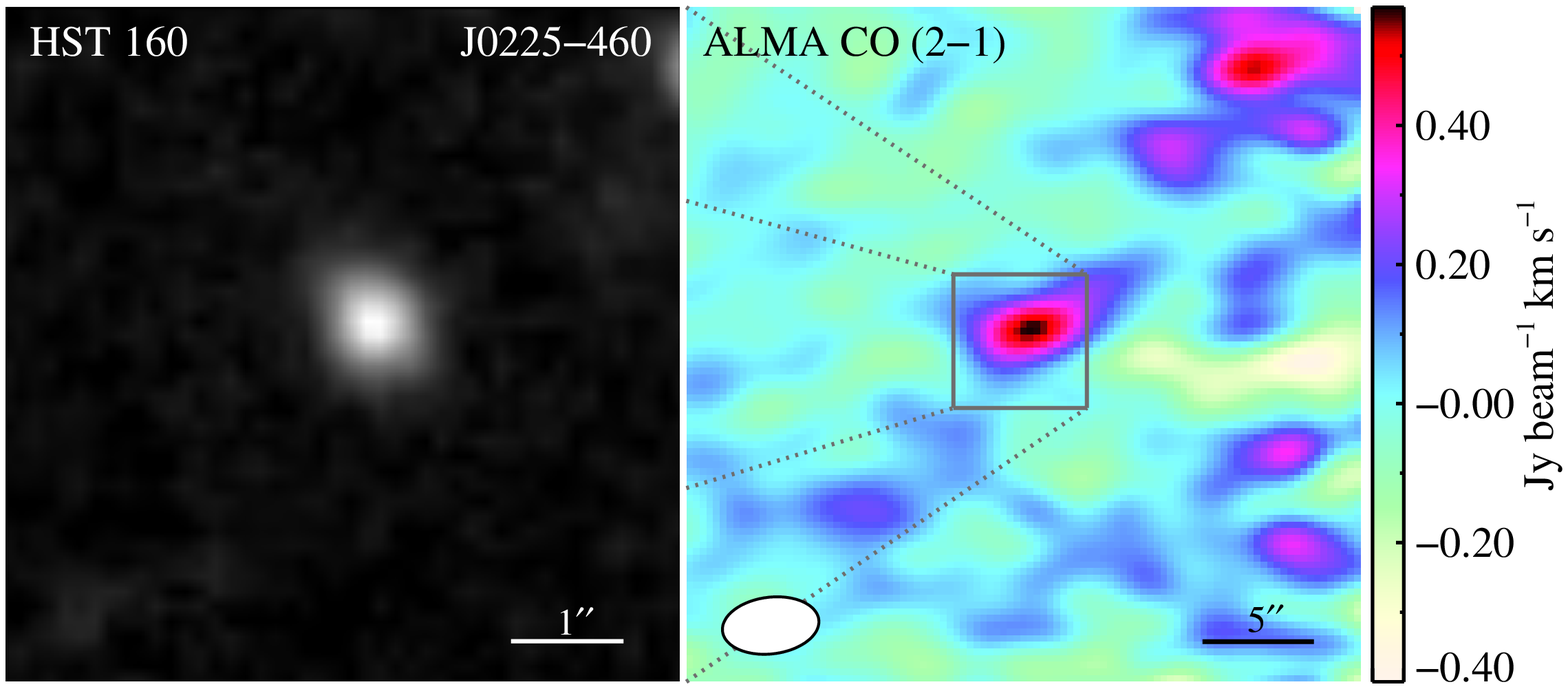}}\hfill%
\subfigure{\includegraphics[width=\subfigsz\columnwidth]{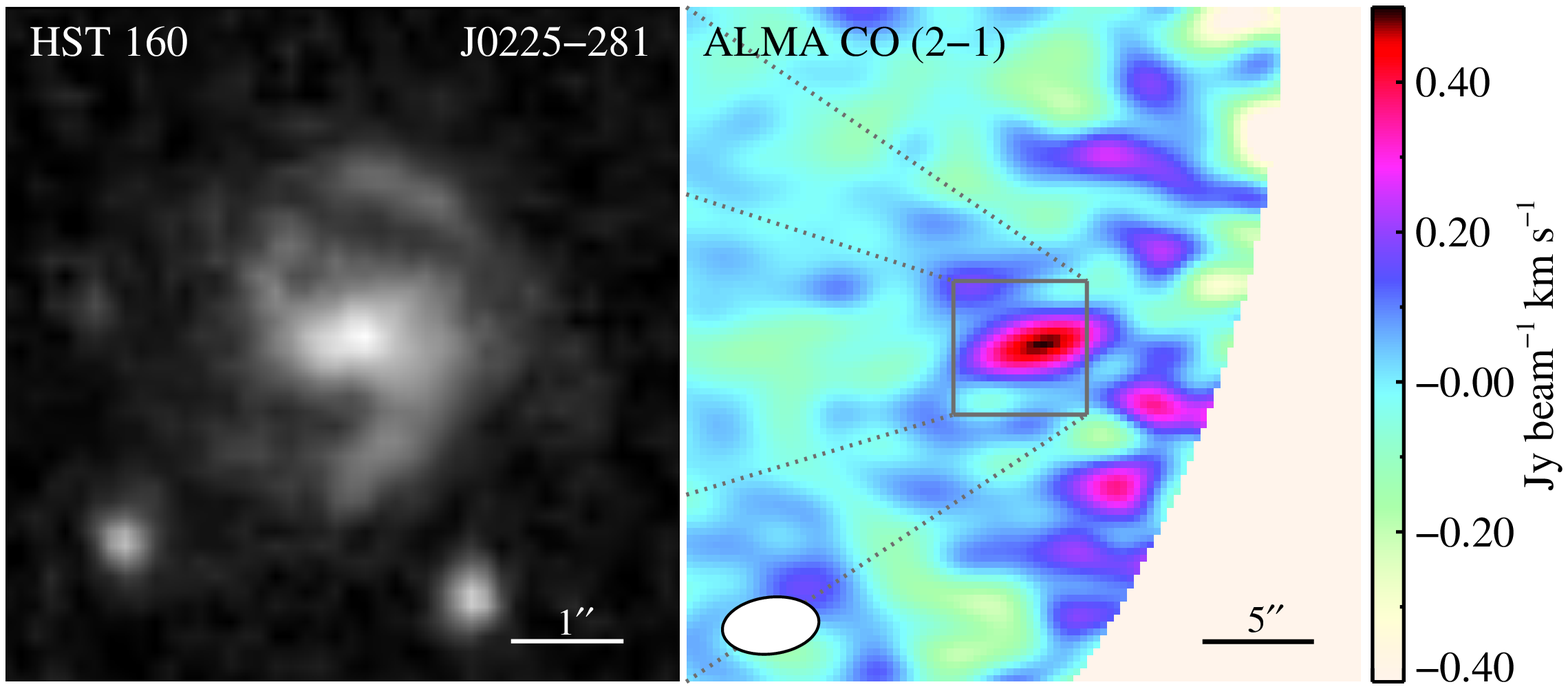}} \hfill%
\subfigure{\includegraphics[width=\subfigsz\columnwidth]{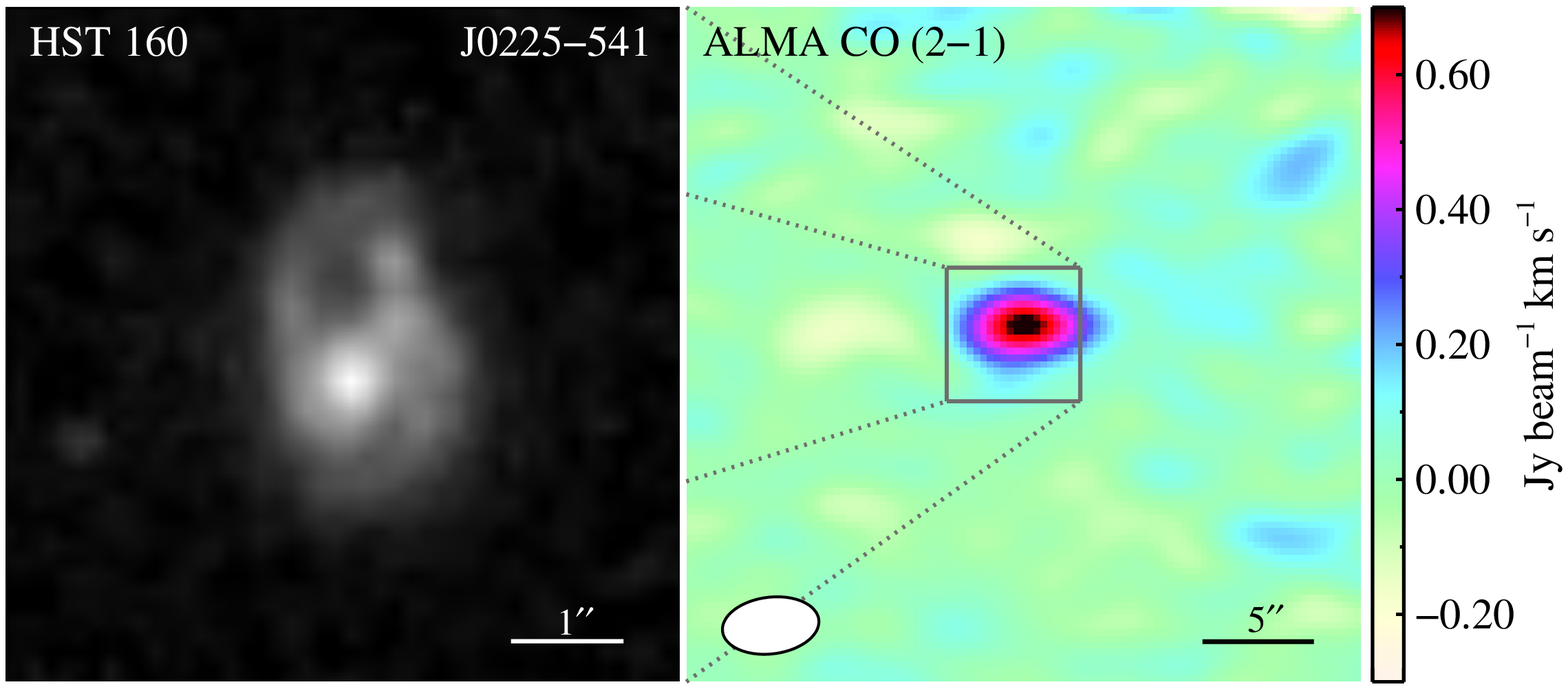}}\hfill%
\subfigure{\includegraphics[width=\subfigsz\columnwidth]{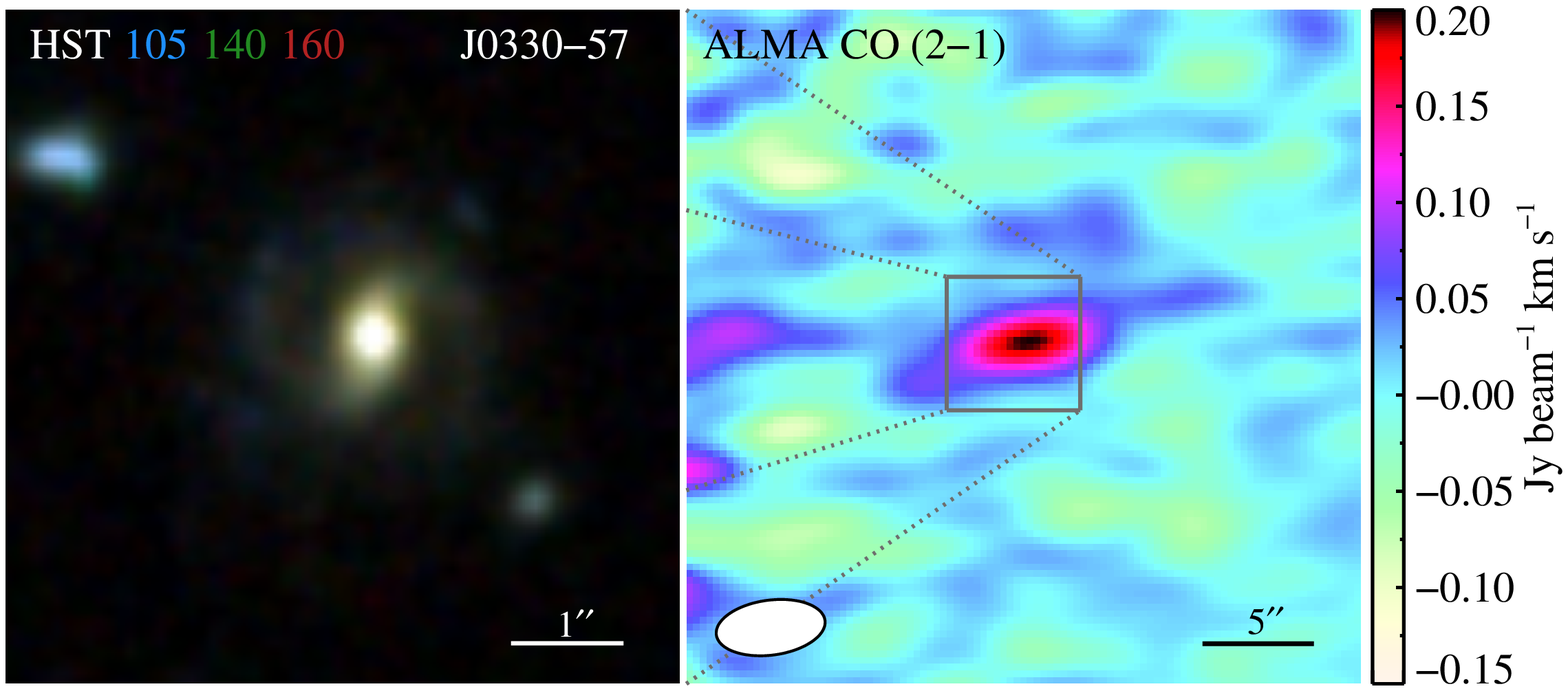}}\hfill%
\subfigure{\includegraphics[width=\subfigsz\columnwidth]{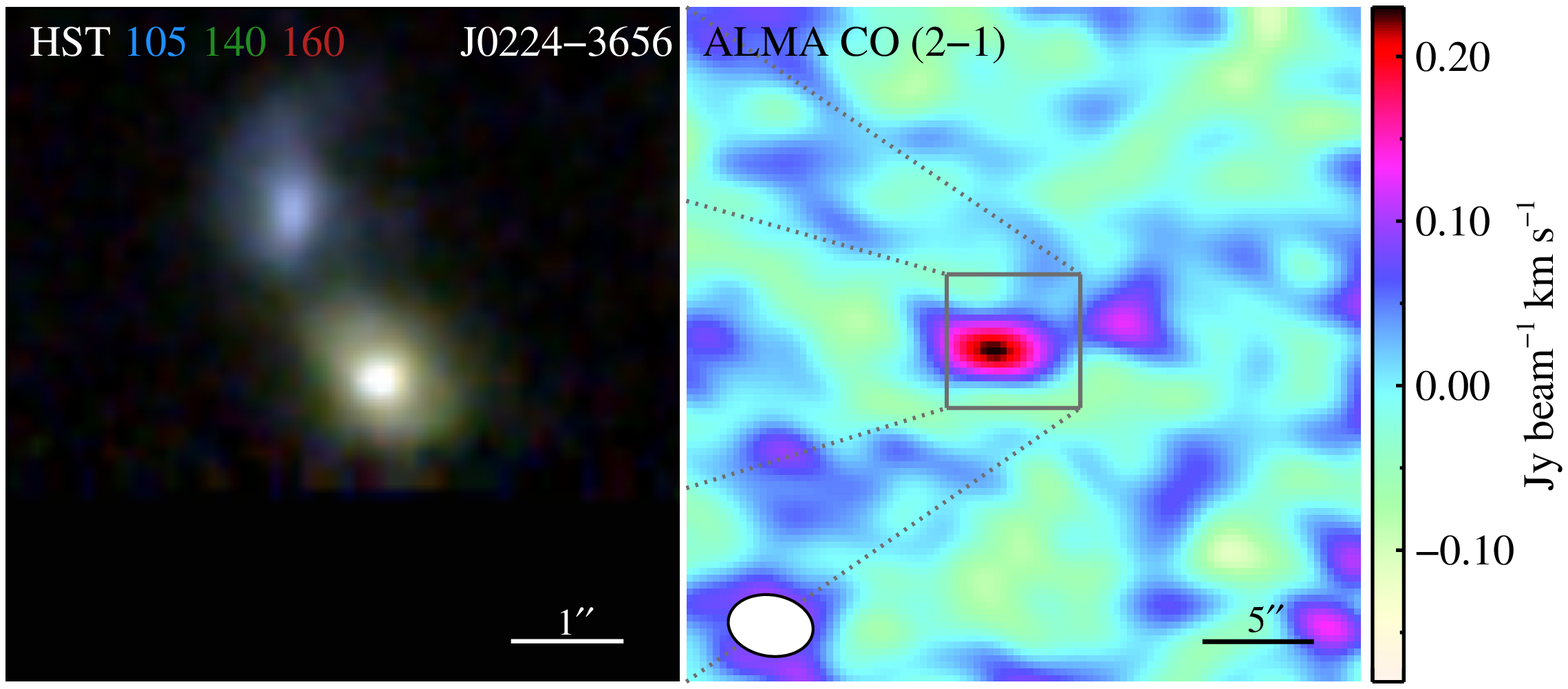}}\hfill%
\subfigure{\includegraphics[width=\subfigsz\columnwidth]{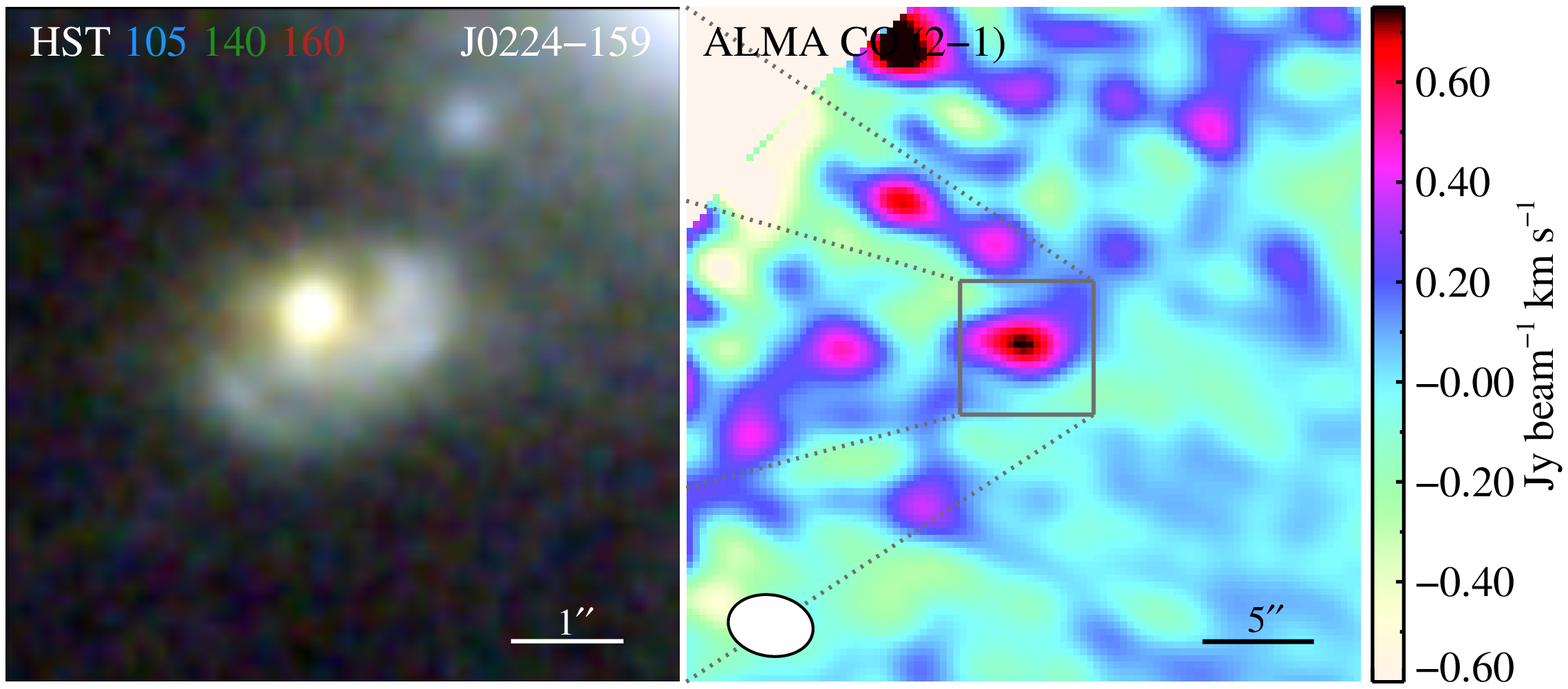}} \hfill%
\subfigure{\includegraphics[width=\subfigsz\columnwidth]{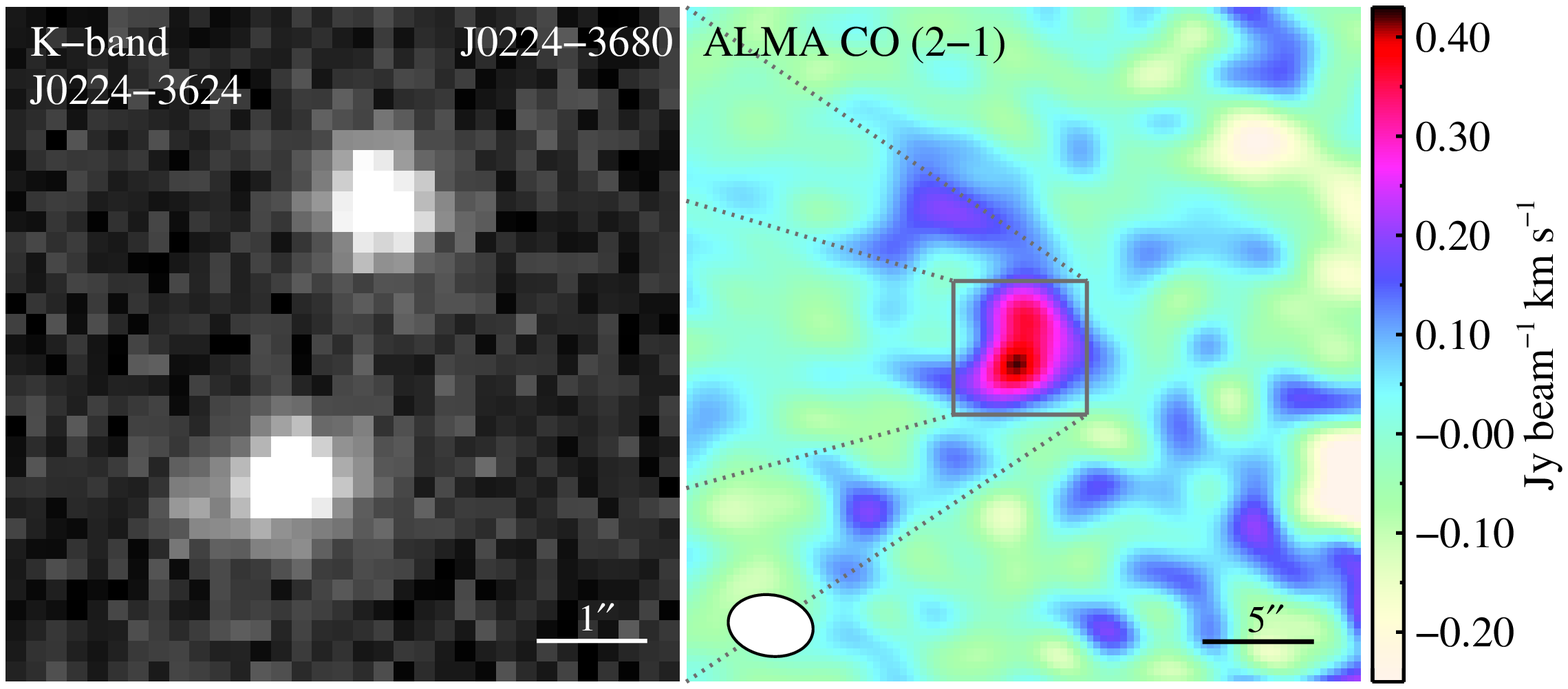}} \hfill%
\subfigure{\includegraphics[width=\subfigsz\columnwidth]{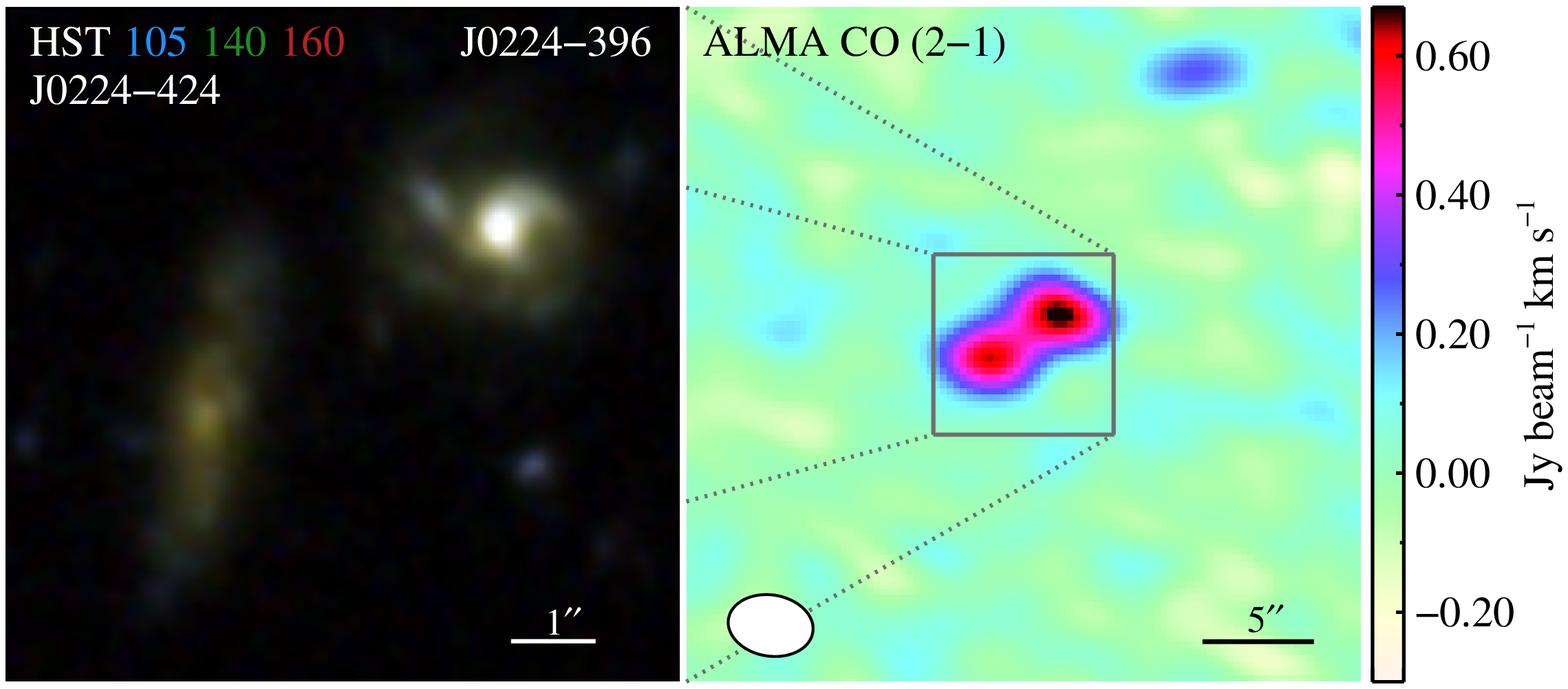}}

\caption{Postage stamps (30\asec$\times$30\asec) showing the CO (2--1) integrated-intensity maps with zoomed-in 6\asec$\times$6\asec\ HST images.  The synthesized ALMA beam for each map is shown by the white ellipse.  We note the blue galaxy to the northeast in J0224--3656 has a photometric redshift of $z=0.63$ and is thus unlikely to be contributing to the CO flux.  The last two stamps in the bottom row represent ALMA pair detections. }
\label{fig:CO}
\end{figure*}

\section{Observations and Analysis}

\subsection{$z\sim1.6$ SpARCS Clusters}

SpARCS J022426--032330 (J0224), J033057--284300 (J0330) and J022546--035517 (J0225) were discovered within the 42 deg$^2$ SpARCS fields \citep[][see Table 1 in Nantais et al. 2016]{Muzzin09, Wilson09, Demarco10}.  They were initially identified using a technique that detects the 1.6\,\um\ stellar bump feature as it spans 3.6 and 4.5\um\ from $1.3<z<1.8$ \citep{Papovich10, Muzzin13}.  These three clusters are spectroscopically confirmed at $z=1.633$, $z=1.626$, and $z=1.59$ \citep{Lidman12,Muzzin13,Nantais16}, respectively, with 115 confirmed members in total.  Richness-based estimates suggest cluster masses $\gtrsim 10^{14}$\,\Msol, placing them amongst the most massive systems at $z\sim1.6$ \citep[e.g.,][]{Stanford12, Bayliss14, Tozzi15, Webb15}.

Additional 11-band imaging exists from optical/near-infrared ($ugrizYK$s) to infrared (3.6/4.5/5.8/8.0\um), allowing for accurate photometric redshifts and stellar masses. Imaging details and analysis are presented in \cite{Nantais16}.  The central cluster regions have deep \text{HST} imaging from the ``See Change" program (GO-13677 and GO-14327) in F160W on the WFC3-IR camera, with additional observations in the F105W and F140W filters for J0224 and J0330. 

\subsection{ALMA Observations}

The ALMA Cycle 3 data were taken between 2016 January 13 and January 20 over 12 execution blocks, with 8.4 hr of total integration time.  Each cluster contains two separate pointings in Band 3, encompassing a total of 49 spectroscopically-confirmed cluster members.  We used the frequency division correlator mode in a single baseband to provide a total bandwidth of 1.875\,GHz. 

The maps were calibrated using ALMA reduction pipeline scripts in CASA version 4.6.0.  We chose 0.\asec 3 pixels and a spectral resolution of 100\,\kms.  We performed minimal cleaning with natural weighting, generating continuum-subtracted and primary-beam-corrected maps with a field of view of $\sim110$\asec\ across.  The resulting data cubes have synthesized beams of $\sim$4\asec$\times$3\asec\ with a central rms of $\sim$0.17\,mJy\,beam$^{-1}$ per channel.

\subsection{CO 2--1 Detections}
\label{sec:CO}

We blindly search the primary-beam corrected image cubes for CO (2--1) detections, requiring a peak S/N $\gtrsim5$, and resulting in a final catalog of 11 CO detections over all 3 clusters.  High-resolution {\it HST} imaging, in conjunction with optical spectroscopy and 11-band photometry, allows for unambiguous counterpart identification of the ALMA detections (Figure \ref{fig:CO}).  Seven of the 11 CO detections represent individual cluster members, and the remaining 4 detections are associated with galaxy pairs.  The ``pair" systems (J0224--3680/3624 and J0224--396/424) are slightly blended in ALMA (and completely blended in the far-infrared imaging).  We therefore treat all the pair detections as single combined systems, measuring total gas masses, stellar masses, and SFRs for each pair.  This yields nine separate flux measurements.

To measure CO fluxes, we first create integrated-intensity map by collapsing the image cube over the velocity channels with significant emission for each source.  We then perform a two-dimensional Gaussian fit on their respective map, and use the best-fit major and minor FWHM to create a $4\sigma$ region for spectral profile extraction on the full image cube.  Within each of these regions, we model the spectral profile with a Gaussian function, determining rms errors from the line-free channels for each source. 

The area under the Gaussian spectral profile corresponds to the full integrated CO flux. These fluxes are subsequently converted into line luminosities (\Lco) using Equation (3) in \cite{Solomon05}. To estimate the total molecular gas mass we use $M_{\rm gas} = \alpha_{\rm CO}(L^{\prime}_{(2-1)}/r_{21})$.  We assume sub-thermalized emission with $r_{21}=0.77$, which is empirically derived in \cite{Daddi15} and consistent with the value used in \cite{Genzel15}, and a \alphaCO\ conversion factor of 4.36 \alphaUnits, commonly used for the Milky Way and in normal star-forming galaxies with solar metallicities \citep{Bolatto13, Genzel15}.  We note that all but one of the cluster galaxies lie within $2\sigma$ of the main sequence of star formation at $z=1.6$.  Table \ref{tab:results} displays our final CO (2--1) measurements, along with corresponding derived quantities.

\begin{deluxetable*}{lccccccccc}
\tablecaption{Properties of the CO-detected cluster galaxies}
\label{tab:results}
\tablehead{
\colhead{ID} &
\colhead{$z_{\rm CO}$} &
\colhead{S/N\tablenotemark{a}} &
\colhead{$S_{\rm CO}\Delta v$\tablenotemark{b}} &
\colhead{FWHM\tablenotemark{b}} &
\colhead{M$_{\rm gas}$\tablenotemark{c}} &
\colhead{M$_{\rm stellar}$} &
\colhead{$\langle\rm SFR\rangle$} &
\colhead{$f_{\rm gas}$} &
\colhead{$\tau$} \\
\colhead{} &
\colhead{} &
\colhead{} &
\colhead{(\Jykms)} &
\colhead{(\kms)} &
\colhead{($10^{10}$\,\Msol)} &
\colhead{($10^{10}$\,\Msol)} &
\colhead{(\myr)} &
\colhead{} &
\colhead{(Gyr)} 
}
\startdata
J0225--371 & 1.599 &  6.3 &  1.18$\pm$0.18 & 401$\pm$71 & 21.9$\pm$3.4 &  6.3$^{+0.8}_{-0.9}$  & 174$\pm$78 & 0.78$^{+0.03}_{-0.04}$  & 1.3$\pm$0.6 \\ 
J0225--460 & 1.601 &  5.8 &  0.63$\pm$0.11 & 509$\pm$104 & 11.7$\pm$2.1 &  9.1$^{+6.0}_{-3.5}$  & 123$\pm$61 & 0.56$^{+0.17}_{-0.10}$  & 0.9$\pm$0.5 \\ 
J0225--281 & 1.610 &  6.2 &  0.59$\pm$0.16 & 122$\pm$34 & 11.1$\pm$3.1 &  6.5$^{+1.7}_{-1.8}$  & 122$\pm$50 & 0.63$^{+0.09}_{-0.09}$  & 0.9$\pm$0.4 \\
J0225--541 & 1.611 & 14.0 &  0.70$\pm$0.06 & 307$\pm$31 & 13.3$\pm$1.2 &  6.6$^{+0.8}_{-0.9}$  &  82$\pm$30 & 0.67$^{+0.03}_{-0.03}$  & 1.6$\pm$0.6 \\
J0330--57 & 1.613 &  5.2 &  0.31$\pm$0.13 & 155$\pm$40 &  5.9$\pm$2.5 &  3.3$^{+1.8}_{-1.5}$  &  36$\pm$21 & 0.64$^{+0.16}_{-0.14}$  & 1.7$\pm$1.2 \\ 
J0224--3656 & 1.626 &  6.8 &  0.30$\pm$0.06 & 539$\pm$113 &  5.8$\pm$1.1 & 10.0$^{+1.2}_{-4.4}$  &  43$\pm$20 & 0.37$^{+0.05}_{-0.11}$  & 1.4$\pm$0.7 \\ 
J0224--159 & 1.635 &  5.2 &  0.46$\pm$0.11 & 245$\pm$68 &  8.9$\pm$2.1 &  5.9$^{+2.6}_{-1.1}$  & 217$\pm$82 & 0.60$^{+0.12}_{-0.07}$  & 0.4$\pm$0.2 \\ 
J0224--3680/3624\tablenotemark{d,e} & 1.626 &  7.0 & 1.07$\pm$0.19 & 776$\pm$192 &  20.5$\pm$3.6 & 9.1$^{+3.5}_{-1.5}$  &  68$\pm$24 & 0.69$^{+0.09}_{-0.05}$  & 3.0$\pm$1.2\\ 
J0224--396/424\tablenotemark{d} & 1.634 &  9.9 &  1.32$\pm$0.12 & 493$\pm$53 & 25.5$\pm$2.4 & 16.2$^{+3.7}_{-2.4}$  & 166$\pm$60 & 0.61$^{+0.06}_{-0.04}$  & 1.5$\pm$0.6 \\ 
\enddata
\tablenotetext{a}{Computed from the peak flux and noise in the collapsed image cube.}
\tablenotetext{b}{Computed from a Gaussian fit to the spectral profile.}
\tablenotetext{c}{Calculated using $r_{21} = 0.77$, $\alpha_{\rm CO}=4.36$.}
\tablenotetext{d}{Pair galaxies, where the CO luminosity and SFR have been measured for the combined system.}
\tablenotetext{e}{Spectral profile is fit with a double Gaussian.}

\end{deluxetable*}

\subsection{Infrared Star Formation Rates}

To obtain SFRs, we utilize infrared/far-infrared data from \textit{Spitzer} and \textit{Herschel}.
All three clusters are within the SWIRE Legacy Survey \citep{Lonsdale03} and the Herschel Mid-infrared Extragalactic Survey \citep{Oliver12}, providing MIPS-24\,\um\ and SPIRE-250/350/500\,\um\ imaging. MIPS counterparts to the ALMA detections are identified directly on the images, and fluxes are measured with aperture photometry. Measurement of SPIRE fluxes is less straightforward due to source confusion in the maps. We attempt to reduce the blending of SPIRE fluxes by using MIPS positional priors and employing a simultaneous stacking technique \citep[SIMSTACK;][]{Viero13}.

As in \citet{Webb15}, we use a Bayesian approach to fit spectral energy distributions to the infrared fluxes.  We first form a two-dimensional parameter space consisting of 105 templates from \cite{Chary01}, each scaled by $10^4$ amplitudes, ranging from 0 to 100. For each template and amplitude combination, we compute the $\chi^2$ value from the observed infrared fluxes, creating a two-dimensional probability distribution.  Assuming flat priors on both the amplitude and template, we calculate the weighted mean over the posterior to determine the infrared luminosity and its uncertainty.  This is converted to a SFR using \cite{Kennicutt98}.  The infrared-derived SFRs place the cluster galaxies around the main sequence at $z=1.6$ from \cite{Whitaker12}.  All but one galaxy (J0224--3656) fall within $2\sigma$ of the main sequence.

\section{Results}
With the first significant sample of CO detections in cluster galaxies at $z\sim1.6$, we can begin to investigate how the cluster environment might impact the molecular gas reservoirs.  We present our main results below.

\subsection{Gas Properties Scaled to the Star-Forming Main Sequence}
\label{sec:offset}

\begin{figure*} \centering
\subfigure{\includegraphics[width=\subfigsztwo\columnwidth]{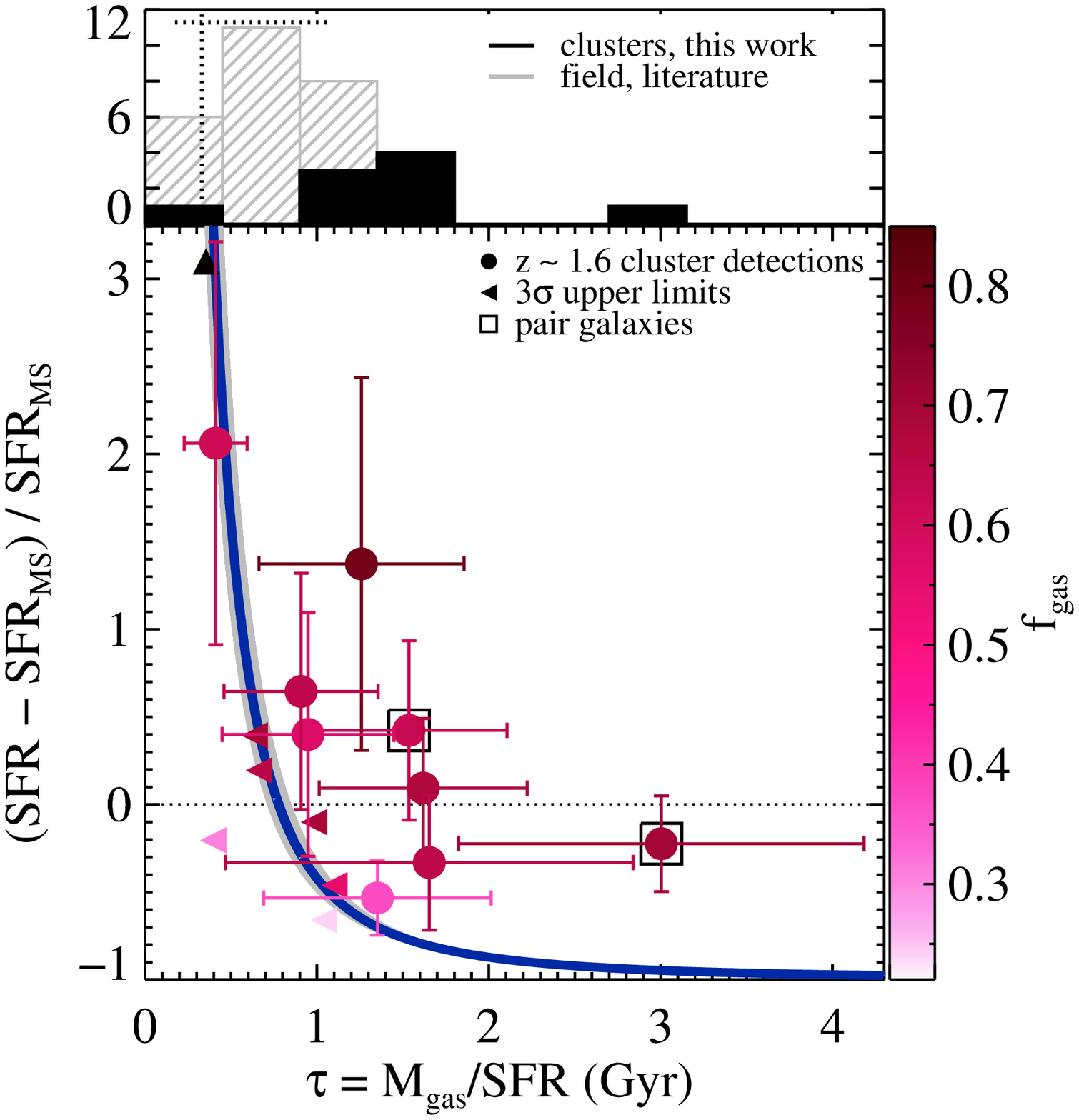}} \hfill
\subfigure{\includegraphics[width=\subfigsztwo\columnwidth]{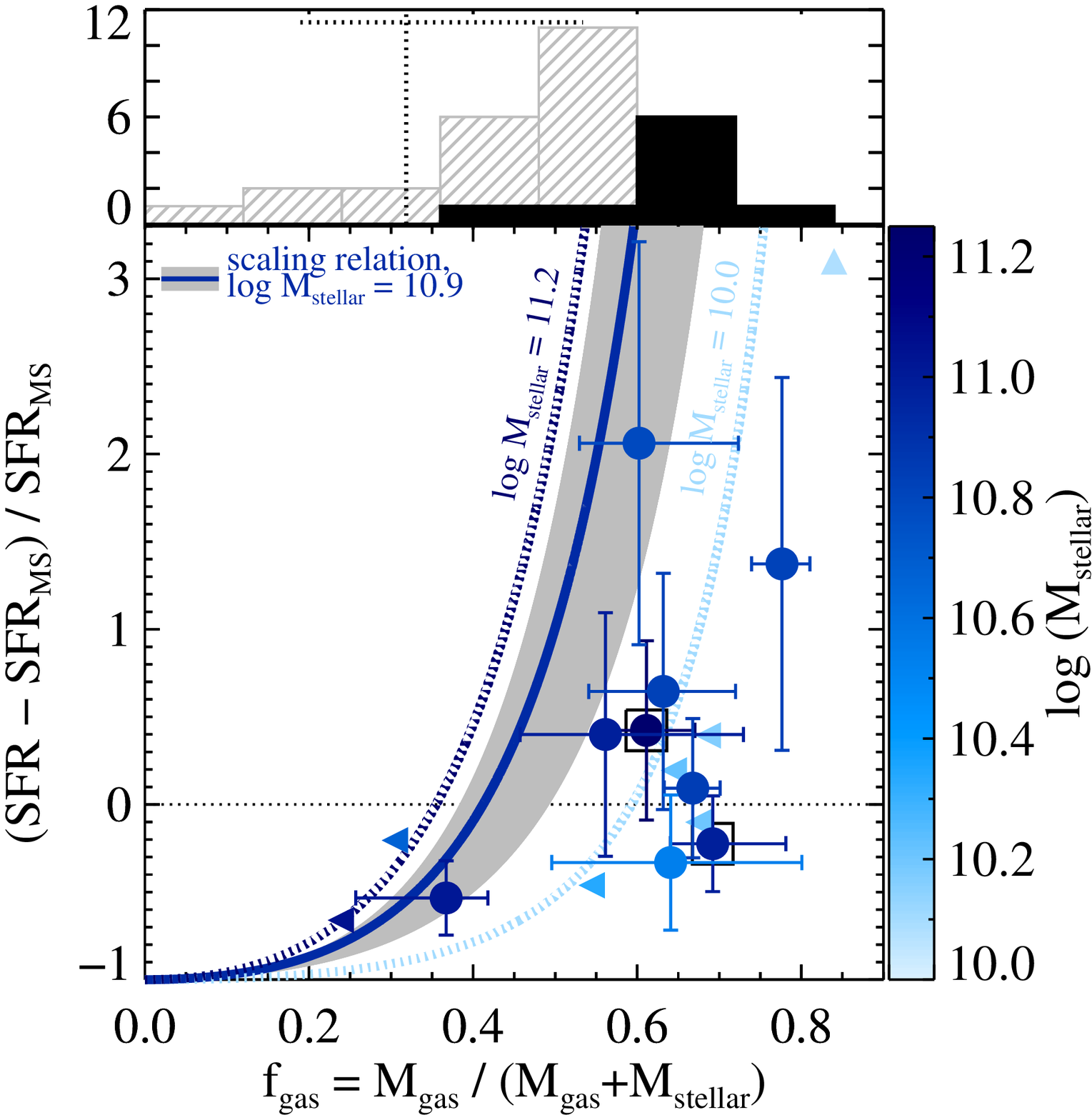}} 
\caption{Relative offset from the star-forming main sequence as a function of molecular gas depletion timescale (left) and gas fraction (right). The $z\sim1.6$ cluster galaxies (circles) are color-coded by the gas fraction (left) and stellar mass (right).  Left-facing triangles correspond to 3$\sigma$ upper limits for spectroscopically-confirmed star-forming cluster members that are not detected in CO (color-coded by their $3\sigma$ upper limit in $f_{\rm gas}$ on the left).  The upward-facing triangle also represents a CO non-detection that has been artificially placed at a lower SFR offset to minimize the plot range; the actual relative offset value is 4.9. The solid blue line (grey region) in both panels represents the field scaling relations ($1\sigma$ fit uncertainties) from \cite{Genzel15}, which have been plotted at $z=1.6$ and normalized to the average mass of our cluster sample.  In the right panel, we also include the scaling relations at the upper and lower mass limits.  The upper panels show the binned distribution of each quantity for our cluster galaxies (filled black) and a similar redshift field sample (lined gray) taken from the literature.  The vertical black dotted lines represent the nominal $3\sigma$ detection limits of our maps, with the horizontal bars depicting the full range of stellar masses and SFRs within our cluster CO-detected sample.  The cluster galaxies lie systematically at higher gas fractions and longer depletion timescales than the field scaling relations.}
 \label{fig:offset} 
\end{figure*}

The tightness of the main sequence of star formation (SFR--$M_{\star}$) is thought to reflect the gas regulator model, in which galaxies grow through an influx of fresh gas that fuels star formation and is subsequently balanced by feedback \citep[][]{Bouche10}. The dependence of gas properties on the galaxy's location on the SFR--$M_{\star}$ plane is therefore expected and has been observed in the field \citep[e.g.,][]{Saintonge11, Saintonge16, Genzel15}.

We investigate the spread of depletion timescales and gas fractions as a function of relative offset from the main sequence  in Figure \ref{fig:offset}.  We show the field scaling relations from \cite{Genzel15}, calculated at $z=1.6$ and normalized to the average stellar mass in our cluster sample.  The gas fraction scaling relation has a steep dependence on stellar mass; we therefore also include tracks using the mass range of the cluster sample.  From these scaling relations, it is evident that field galaxies further above the main sequence display higher gas fractions and shorter depletion timescales.  

For a given mass, the $z\sim1.6$ cluster galaxies lie at systematically higher gas fractions than the scaling relation (of the appropriate mass). The same is true of depletion timescales, though most are within a one standard deviation of the relation.   We quantify this offset by summing the individual $\chi^2$ values from each data point compared to model track scaling relation at the cluster galaxy's given stellar mass.  This returns a  $8\times 10^{-7}$ ($0.3$) likelihood of producing a similar or worse $\chi^2$, roughly corresponding to a  $\sim5\sigma$ ($1\sigma$) offset in gas fractions (depletion timescales) compared to the field scaling relations. 

We explore whether this could be due to a selection effect by including $3\sigma$ upper limits for spectroscopically confirmed infrared-detected cluster members above the scaling relation mass limit of $10^{10}$\,\Msol.  For each of these seven non-detections, we create a 400\,\kms\ (the average FWHM of the detected sample) integrated-intensity map centered at the spectroscopic redshift. The pixel-to-pixel variation in an annulus around the source corresponds to the $1\sigma$ rms.  We then estimate the $3\sigma$ upper limit on the gas fraction and depletion timescale using the galaxy's stellar mass and SFR.  
Most of the non-detections lie close to the scaling relations, albeit fewer than the number of CO-detected galaxies lying above.  Therefore, while we cannot rule out the existence of a cluster population consistent with the field, there is still a higher fraction of cluster galaxies offset from the relation given the uncertainty in the scaling relation fit.  We can thus reasonably rule out that the offset is purely a selection effect.  Moreover, if we include the upper limits in the $\chi^2$ calculation by conservatively assuming the non-detections lie on the scaling relations, the offset significance for the gas fraction only drops to  $\sim4\sigma$.  We note that there is likely some selection bias in the field samples, further muddling interpretation.  

We compare the distribution of  gas properties to coeval field galaxies from  $1.2<z<1.6$ \citep{Daddi10, Tacconi13, Decarli16, Papovich16} in the upper panel histograms.  We restrict the field sample to galaxies within a similar range of offsets from the main sequence as the cluster CO-detected sample, from $-1$ to $2$. We note the cluster and field comparison samples are evenly distributed on the SFR--M$_{\star}$ plane.  The tendency toward higher gas fractions in cluster galaxies is again conspicuous.  To evaluate the differences, we restrict the analogous field sample to values of $f_{\rm gas}$ and $\tau$ above our nominal $3\sigma$ detection limit in the cluster sample.  This is estimated using the typical rms in the center of the ALMA maps and the average FWHM, stellar mass, and SFR of our detected sample, yielding a gas fraction and depletion timescale limit of 32\% and 0.33\,Gyr.  Comparing the two distributions above our nominal detection limits, we perform a Kolmogorov--Smirnov test, rejecting the null hypotheses in both cases with 99\% confidence.   We find an average gas fraction of $62\pm3.7\%$ and average depletion timescale of $1.4\pm0.2\,$Gyr for our CO-detected cluster galaxies.

\subsection{Evolution of the Gas Fraction in Clusters}

\begin{figure} \centering
\includegraphics[width=1.0\columnwidth]{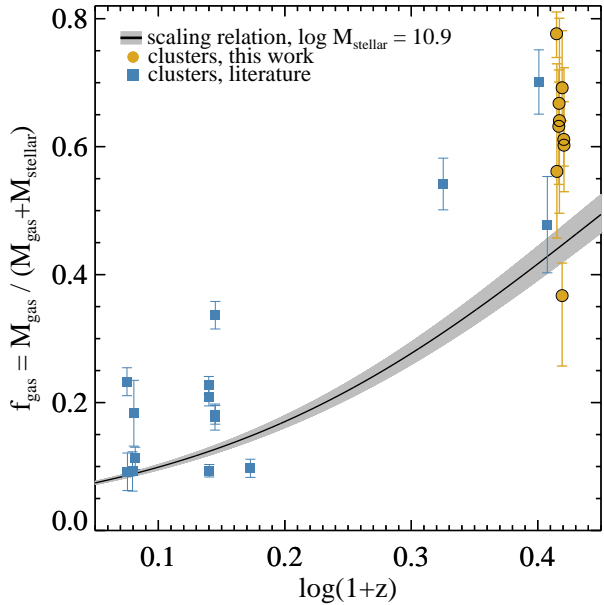}\hfill
\caption{Evolution of the gas fraction for main-sequence cluster galaxies, compared to the field scaling relation \citep[black line and gray region,][]{Genzel15}.  The lower-redshift cluster data are taken from the literature and limited to a narrow range around the main sequence at their respective redshifts.  On average, $z>1$ main-sequence cluster galaxies have higher gas fractions than the coeval field. }
\label{fig:evol}
\end{figure}

In Figure \ref{fig:evol}, we plot the evolution of the gas fraction. We compile a subset of 19 additional CO detections in clusters from the literature from $0.2<z<1.5$ \citep{Geach11, Aravena12, Wagg12, Jablonka13,  Cybulski16} to compare to our $z\sim1.6$ detections.  We similarly restrict the literature detections to $>10^{10}$\,\Msol\ galaxies that fall within a relative offset from $-1$ to $2$ of the main sequence at their respective redshift, yielding 15 cluster galaxies.  Including galaxies markedly above the main sequence, for example, would inherently bias the literature detections to higher gas fractions given the aforementioned correlation in \S\ref{sec:offset}.  We include the rise in the gas fraction for main-sequence field galaxies from the \cite{Genzel15} scaling relations, normalized to the average mass of our cluster galaxies.  The gas fraction in cluster galaxies mimics the strong evolution in the field.  Notably, almost all the $z>1$ cluster galaxies lie above the gas fractions in main-sequence field galaxies, despite half the galaxies lying slightly below the main sequence.  Conversely, gas-rich galaxies in low-redshift clusters are on average closer to the field gas fractions.  This is suggestive of a steeper evolution in gas fractions for cluster galaxies than the field, consistent with semi-analytical \citep{Lagos11}  and semi-empirical \citep{Popping15} models that predict a stronger evolution in more massive halos.  However, this warrants caution owing to the heterogeneous nature of the cluster and field samples, making interpretation difficult.  For example, this could  be dominated by systematic offsets in SFR measurements and/or selection biases.

\section{Discussion}
We find that star-forming main-sequence cluster galaxies are systematically concentrated toward higher gas fractions compared to the field scaling relations at $z\sim1.6$.

This could partially be a selection effect---we cannot fully exclude the possibility that with deeper data we would detect more galaxies on or below the scaling relations.  However, it is unlikely the sole cause of the offset as more of our confirmed cluster members have CO detections as opposed to non-detections consistent with the scaling relations.  Barring a selection effect, we propose three other plausible explanations for the offset in gas properties of cluster galaxies relative to the field.

One possibility could be that for a given gas fraction, SFRs in $z\sim1.6$ cluster galaxies are suppressed.  While the SFRs in CO-detected cluster galaxies range within $\sim2\sigma$ of the main sequence, one might expect even higher levels of star formation given the massive gas reservoirs.  Though it seems unlikely that star formation would begin to cease before the depletion of molecular gas \citep{Bahe15}, this could be due to the varying timescales of the measurements \citep{Feldmann17}.

Conversely, taken at face value, this offset implies that star-forming main-sequence galaxies in cluster environments have higher gas masses than the field.  This could be suggestive of an environmental interaction that perturbs the molecular gas in cluster galaxies such that a smaller fraction of the gas actively contributes to star formation; this would require higher gas masses in cluster galaxies compared to field galaxies with the same SFR.  Similarly, environmental pressure could increase the formation of molecular gas through compression of the interstellar medium that further prevents gaseous outflows \citep{Fujita99, Bahe12}, yielding higher gas masses than field galaxies for a given stellar mass.   Indeed, simulations find an increased effectiveness of ram pressure at $z\sim1$ compared to $z=0$ \citep{Bahe15}.  Moreover, Virgo cluster galaxies have also been found to have an excess of molecular gas despite being deficient in H\textsc{i} \citep{Mok16}, though this is in contrast to many other studies that report a reduction of molecular gas \citep[e.g.,][]{Boselli14}. Large gas reservoirs in $z>1.5$ clusters are also consistent with the increased star formation observed in dense regions at this epoch \citep[e.g.,][]{Tran10}.

Finally, the same \alphaCO\ may not be appropriate for field and cluster galaxies alike.  We note that reducing the \alphaCO\ conversion by $2\times$ for the cluster galaxies would remove the offset between the sample and the field scaling relations.  This term is dependent on various factors, most notably metallicity and total mass surface density \citep{Bolatto13}, both of which could be affected by the larger-scale environment.  Indeed, the value of \alphaCO\ slightly decreases for increasing metallicity \citep{Narayanan12}.  Although galaxies in high-density environments have marginally higher metallicities \citep{Cooper08b}, a factor of $\sim2$ increase in metallicity would be needed to reduce \alphaCO\ in our cluster galaxies in order to align them with field gas fractions.
A lower value of \alphaCO\ is also preferred for mergers (\alphaCO\ $\sim1$), due to a combination of increased gas temperatures and velocity dispersions that give rise to an amplified CO luminosity \citep{Narayanan12}.  While we do see examples of pair galaxies in our cluster sample, recent work by \cite{Delahaye17} finds no direct evidence for increased merger activity in $z\sim1.6$ cluster cores compared to the field.     In addition to mergers, a similar effect could also result from ram-pressure stripping, where compressed gas at the leading edge of the galaxy would lead to higher gas temperatures and velocity dispersions, necessitating a lower conversion between CO and H$_2$.   If cluster galaxies indeed warrant a different \alphaCO\, this in itself is interesting as it implies that environmental studies of molecular gas need to be more cognizant of systematic \alphaCO\ variations.

\section{Conclusion}

We present the largest study of molecular gas in $z>1.5$ cluster galaxies to date.  Using ALMA Band 3, we detect CO (2--1) in 11 galaxies over 3 massive SpARCS galaxy clusters.  We summarize our results as follows:

\begin{enumerate}

\item{The $z\sim1.6$ cluster galaxies have consistent depletion timescales ($\bar{\tau}=1.4\pm0.2\,$Gyr), but $\sim4\sigma$ higher gas fractions ($\bar{f}_{\rm gas} = 62\pm3.7\%$) for a given offset from the main sequence compared to the scaling relations of coeval field galaxies.}

\item{Cluster galaxies on or around the main sequence mimic the strong evolution in the gas fraction in the field, with the trend continuing in clusters up to $z\sim1.6$.}

\end{enumerate}

The origin of the gas fraction excess is not clear---whether it is a selection effect or stems from a cluster environmental dependency remains an open question.
Larger samples of high-redshift CO detections in cluster galaxies are required, preferentially probing a broad scope in the SFR--$M_{\star}$ plane, and over a wide range of cluster halos to mitigate stochastic cluster-to-cluster variations.  Nevertheless, with these data, it is clear that high-redshift galaxy clusters have an ample supply of gas-rich galaxies.  Given the efficiency of targeting high-density fields to obtain multiple detections within a single field of view, clusters offer an exciting laboratory to further explore molecular gas properties.

\acknowledgments
We thank the referee for providing feedback that improved this Letter, and H. Russell for useful discussions. We acknowledge support from the following sources: NSF grants AST-1517815/AST-1211358/AST-1517863/AST-1518257; NASA programs AR-14310.001/GO-12945.001-A/GO-13306/GO-13677/GO-13747; Universidad Andres Bello grant DI-18-17/RG; FONDECYT 1130528; BASAL CATA; ISSI.

This Letter makes use of the following ALMA data: ADS/JAO.ALMA\#2015.1.01151.S. ALMA is a partnership of ESO (representing its member states), NSF (USA) and NINS (Japan), together with NRC (Canada) and NSC and ASIAA (Taiwan) and KASI (Republic of Korea), in cooperation with the Republic of Chile. The Joint ALMA Observatory is operated by ESO, AUI/NRAO and NAOJ.  The NRAO is a facility of the NSF operated under cooperative agreement by Associated Universities, Inc.
Based on observations made with the NASA/ESA \textit{HST} (GO-13677/GO-14327), obtained at the STSI, which is operated by the AURA, Inc., NASA contract NAS 5-26555. 

%

%

\end{document}